%% file: paper.tex
\documentclass[12pt]{iopart}

\usepackage{graphicx,a4,pslatex,cite,subfigure,amsfonts}

\begin{document}

\input{title_abstract.tex}

\input{intro.tex}

\input{Model.tex}
\input{quadtree.tex}

\input{Analysis.tex}
\input{Results.tex}
\input{Conclusion.tex}

\section*{References}
\bibliographystyle{unsrt}
\bibliography{bib/Phys_Rev_E.bib,bib/Phys_Rev_Lett.bib,bib/uralt.bib,bib/Physica_A.bib,bib/Eur_Phys_J_E.bib,bib/sonstige.bib,bib/condmat.bib}

\end{document}

%% file: title_abstract.tex
\title{Shearing behavior of polydisperse media}

\author{Martin Wackenhut, Sean McNamara, Hans Herrmann
}
\address{Institute for Computational Physics, University of
Stuttgart,    
Pfaffenwaldring 27, 70569 Stuttgart, Germany}
%  \and \\ICA1, Universität Stuttgart\\
%Pfaffenwaldring 27, 70569 Stuttgart, Germany}
\ead{M.Wackenhut@ica1.uni-stuttgart.de}
\ead{S.McNamara@ica1.uni-stuttgart.de}
\ead{H.J.Herrmann@ica1.uni-stuttgart.de}

%\pacs{45.70.-n, 82.20.Wt, 81.05.Rm}
%\maketitle

\begin{abstract}
We study the shearing of polydisperse and bidisperse media with a size ratio of 
10. Simulations are performed with a the two dimensional shear cell
using contact dynamics. With a truncated power law for the polydisperse media
we find that they show a stronger dilatancy and greater resistance to shearing
than bidisperse mixtures. 
Motivated by the practical problem of reducing the energy needed to shear granular
media, we introduce "point-like particles" representing
charged particles in the distribution. Even though changing the kinematic behavior 
very little, they reduce the force necessary to maintain a fixed shearing velocity.
\end{abstract}

%% file: intro.tex
\section{Introduction}

	Granular media show a wide variety of phenomena and people tried to understand
	these phenomena using continuum mechanics.
	Parallel to that, computer models like Molecular Dynamics \cite{Phys_Rev_E_55_4_4720}
	or contact dynamics \cite{Phys_Rev_E_54_1_861, Phys_Rev_Lett_77_2_274,Moreau} are
	used to simulate granular media on the individual grain level.  The
	application of external stress to a granular system leads to force chains.
	Along these chains the system carries the majority of the stress while there
	are regions with small or no stress.  When the system is sheared, these
	chains break up and stress fluctuations can be observed
	\cite{Phys_Rev_Lett_77_15_3110,Chaos_9_3_559}.

	Dilatancy is one of the fundamental properties of granular media and first
	studies where done by Reynolds in 1885 \cite{Philos_Mag_20_469}.  Special
	interest lies in understanding dilatancy due to the shearing of granular
	materials. Here experiments on granular media immersed in water,
	\cite{Phys_Rev_E_59_5_5881}, sheared granular layers,
	\cite{Phys_Rev_Lett_79_5_949} and two-dimensional granular Couette
	experiments \cite{Phys_Rev_E_59_1_739} contributed many insights. On the
	other hand computational physics helps to study these complex systems.
	Thompson and Grest \cite{Phys_Rev_Lett_67_13_1751} use molecular dynamics on
	disks while Tillemans and Herrmann \cite{Physica_A_217_261} use polygons.
	Other numerical work was done by Lacombe et al.  \cite{Eur_Phys_J_E_2_2_181}
	and Lätzel \textit{et. al.} \cite{Eur_Phys_J_E_2002_10160_7}.
	
%%	Schollmann \cite{Phys_Rev_E_59_1_889}.  
	Recent work \cite{condmat_0407100} compares results given by the Enskog equation and results from a
	MD for polydisperse granular fluids under shear.
	In the present studies we will
	focus on the shearing behavior of polydisperse mixtures.  Therefore we 
	compare it to a bidisperse
	mixture and study the changes subject to the introduction of point
	particles.

%% file: Model.tex
\section{The model}
	\label{sec:model}
	First we will define different particle distributions and introduce our
	"point-like particles". Afterwards we give an overview over the simulation
	method including the procedure we used to initialize the system.  Finally
	compaction and shearing of the system is explained.

	\subsection{Grains}
		We simulate a two-dimensional system of circular disks which we
		will refer to as "grains".
		We define the polydispersity $\Pi$ of our mixtures by:
		\begin{equation}
			\Pi=\frac{r_{max}}{r_{min}}
		\end{equation}

		A bidisperse distribution contains particles of two different sizes. The
		radius of the large particles be $r_{max}$ while the radius
		of the small particles be $r_{min}$.  To fully define this
		distribution we introduce the ratio $R$ between the number of big
		particles and the number of small particles by:
		\begin{equation}
			R= \frac{N(r_{max})}{N(r_{min})}
			\label{eq:bidispers}
		\end{equation}
		Here $N(r)$ is the number of particles of a given size $r$. 
	
		The polydisperse distribution is given by a truncated power law:
		\begin{equation}
			P(r)=a\cdot r^{-b}
			\label{eq:powerlaw}
		\end{equation}
		Here $r$ is the radius of the
		particle, $b$ the exponent of the power law and $a$ a prefactor which is
		chosen such that $\int_{r_{min}}^{r_{max}} P(r)dr =1$, where $r_{min}$
		and $r_{max}$ are the minimum and maximum radius in the distribution. 

		The mass $m$ of the grains is given by:
		\begin{equation}
			m=\frac{4}{3}\pi r^3
		\end{equation}

			For resolving the grain-grain interactions we use contact dynamics 
			where we have set the restitution coefficient $r_c$ to $0.2$.
			Additionally we use Coulumb friction, where the Coulomb force 
			is defined by: 
			\begin{equation}
				F_C=\mu	F_n
				\label{eq:coulomb} 
			\end{equation}
	
			We determine the tangential force $F_T$ necessary to reduce the
			tangential velocity $v_t$ of the contact to zero. In the case of 
			sliding friction, $F_T$ is	larger than $F_C$ and we reset it to
			$F_C$ before applying it to the contact. 
			In the case of static friction, we apply $F_T$
			as it is smaller than $F_C$.
		
		\subsection{Point-like particles}
		
%			\bild{repulsive.eps}{Picture of a repulsive particle}{fig:repulsive}
%			\noindent
			A point-like particle has a zero radius, zero mass and interacts with 
			grains through the potential:

			\begin{equation}
					U(d)= \left\{ \begin{array}{ll}
					kd_r \hat d^{-1}e^{-\alpha \hat d}
						+F_0 \hat d+U_0, & 0<\hat d<1\\
					0, & \hat d\ge 1
												\end{array} \right. 
				\label{eq:repulsivepotential}
			\end{equation}
			Here $d_r$ is the interaction radius of the point-like particle
			and $\hat d=d/d_r$ is the dimensionless distance between the particle
			surfaces, where $d$ is the separation
			between the particle surfaces. Note that when $d>d_r$ the particles do not
			interact.
			
			The first term in Eq. \ref{eq:repulsivepotential} 
			is a screened long-range potential.
			This term contains two constants:
			$k$ determines the strength of the potential while
			$\alpha$ fixes how fast it decays with the distance.
			In this work, we fix $\alpha=3$ and adjust the range of the
			potential through $d_r$.
			The second and third terms are small, and added for numerical convenience.
			The constants $U_0$ and $F_0$ are chosen such that potential and force
			are continuous at $d=d_r$. 

			The repulsive force $F_r$ is just the gradient of the potential:
			\begin{equation}
				F_r(d)=-\frac{\partial U}{\partial d}= \left\{ \begin{array}{ll}
				k\hat d^{-1} e^{-\alpha\hat d}(\hat d^{-1}+\alpha)-F_0 & 0<\hat d<1 \\
												0 & \hat d\ge 1
												\end{array} \right. 
				\label{eq:repulsiveforce}
			\end{equation}
			
			Point-like
			particles don't contribute to the density of the system but exert a
			repulsive force $F_r$ on every particle closer than the distance
			$r_r$.
			As they have no mass and thus might experience
			infinite accelerations when using contact dynamics we
			use a different iteration scheme for calculating their motion.
			Before beginning a contact dynamics time step, each
			point-like particle is moved to a position where the net force on it
			vanishes. Then the forces exerted by the point-like particles on the
			other particles are computed, and the contact dynamics time steps proceeds
			normally.

	\subsection{Simulation Method}
		In this section we will explain the setup we used to investigate the shear behavior 
		of particle mixtures immersed in a fluid. 
		First we fill the	shear cell with an initial configuration of particles
		then we compact this configuration
		and in the last stage we shear the system.

		\subsubsection{The shear cell}
			\begin{figure}
			\centering
			\includegraphics[width=0.7\textwidth]{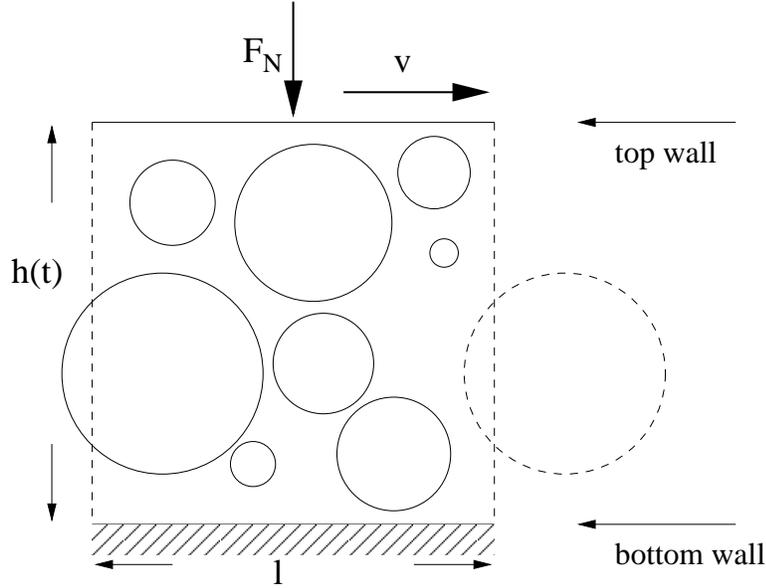}
			\caption{The shear cell with periodic boundaries.
			The length of the cell is $l$ and the actual hight of the lid is given
			by $h(t)$. While we shear the lid with the velocity $v$ we exert
			the normal force $F_N$ onto the lid.
			}
			\label{fig:model}
			\end{figure}
			\noindent
			
			Fig.\ref{fig:model} shows the two dimensional shear cell.  We apply
			periodic boundary conditions such that particles leaving the cell to
			the left will reenter on the right and vice versa. The length of the
			system is $l=10cm$ and we set the density of the particles to
			$\rho=10^3kg/m^3$.  The bottom wall is fixed while on the top wall,
			the lid, we exert a force $F_N=100 N$ in all simulations. The position
			of the lid is given by its height $h(t)$.  After compaction we shear
			the lid with the velocity $v$.
		
			In all simulations we turned off gravity.
			
		\subsubsection{Initialization}
			In order to obtain a high starting density we use a hierarchical
			initialization scheme. First we fill a separate reservoir with 
			either the bidisperse or the polydisperse mixture.
			While filling the	reservoir we calculate the area $V$ of all particles
			in the reservoir.
			\begin{equation}
				V=\sum_i^n\pi r_i^2
				\label{eq:filling_volume} 
			\end{equation}
			Here $r_i$ is the radius
			of the $i$th particle and $n$ is the number of particles in the
			reservoir. When $V$ becomes equal or greater than 70\% of the shear cell volume
			we stop filling the reservoir.
			Next we change the indices of the $n$ particles in the reservoir 
			 such that $r_1 >	r_2 > ... > r_n$. Starting with the
			largest particle we consecutively take the next smaller particle
			and give it $I$ trials to find a random position in the shear
			cell where it does not overlap with any particle already present or the
			top and bottom wall. The value for $I$ is chosen high enough that 
			all $n$ particles can be placed in the shear cell.
			
			If point-like particles are used in the simulation they are put in next.
			We take $N_p$ point-like particles and give them $I_p$ trials for adsorption.
			As	$r=0$ all $N_p$ point-like particles can be put into the system
			for a large enough $I_p$. During the initialization the strength of
			the potential is always set to $k_i=10^{-3} N$ and before we start
			shearing, $k$ is set to the wanted value. 
			This makes it faster to reach our desired initial density when compressing
			as $k$ is greater of equal to $k_i$ in all our simulations.
			
			When all particles have been placed in the shear cell, the
			initialization is complete. 
			
			When averaging over several simulation we always use the same $n$
			grains and the same number $N_p$ of point-like particles in the
			reservoir but use a different seed for the random number generator
			responsible for choosing the positions of the particles.  Thus the
			distribution of particle sizes remains exactly the same, only
			the initial	configurations are different.

		\subsubsection{Compaction}
			The actual volume fraction $\Phi$ of the system is given by:
			\begin{equation}
				\Phi(t)=\frac{V}{lh(t)}
				\label{eq:density}
			\end{equation}
			Starting from the
			initial configuration we want to reach a specific volume fraction
			$\Phi_0$ at which we start the shearing. To reach $\Phi_0$ we exert a
			fixed normal force $F_N$ on the lid and, in certain time intervals, we give the
			grains a random force in a random direction.  Due to the force $F_N$
			the system is compressed while the random forces break up arches and
			thus allow for better compaction.  Additionally we turn off friction
			($\mu=0$).  As soon as the
			volume fraction $\Phi$ reaches the desired value $\Phi_0$ we stop the
			compaction, turn on friction ($\mu=0.3$) and start shearing.

		\subsubsection{Shearing}
			In the compacted system we determine the grains with a radius
			smaller than $2r_{min}$ and a distance smaller then $r_{min}/2$
			away from the top and bottom wall and fix them to these walls. The
			position of the lid at the moment we start to shear is the reference
			height $h_0$. As we shear the lid with a constant velocity $v$, we
			measure the position $h$ of the lid and the force $F$ we need to exert
			on it to keep $v$ constant. 
			We shear for at least 0.1 seconds
			in all simulations. For the slowest shear velocity the lid moves at
			least a distance $l/2$ in horizonal direction.

%%		\subsubsection{Quad-Tree}

%% file: quadtree.tex
\subsubsection{Particle contact detection}%%The Quadtree}
	
	Now we will consider the detection of the particle contacts.  A well know
	method is the "linked cell" algorithm \cite{Tildesley}.  Here a grid of
	equal sized cells of side length $s$, the radius
	of the biggest particle, is put over the system and each particles is 
	assigned to the cell in which its center lies in. In the next step the "Verlet
	list", containing pairs of particles whose separation is smaller than a
	certain threshold $d_t=s/2$, is created by determining 
	contacts between the particles in one cell and those in the
	neighboring cells. With neighboring cells we refer to the nearest and 
	next nearest neighbors. Thus a cell has eight neighbors in two dimensions.
	When creating the Verlet list it is sufficient to check only half of the
	neighbouring cells.
	If any particle travels further than $d_t$ the Verlet list is regenerated.
	When using this method for very polydisperse media, each cell will 
	contain many small particles which will slow down the
	simulation as we put many unnecessary contacts into the Verlet list.

	\begin{figure}
		\centering
		\includegraphics[width=0.7\textwidth]{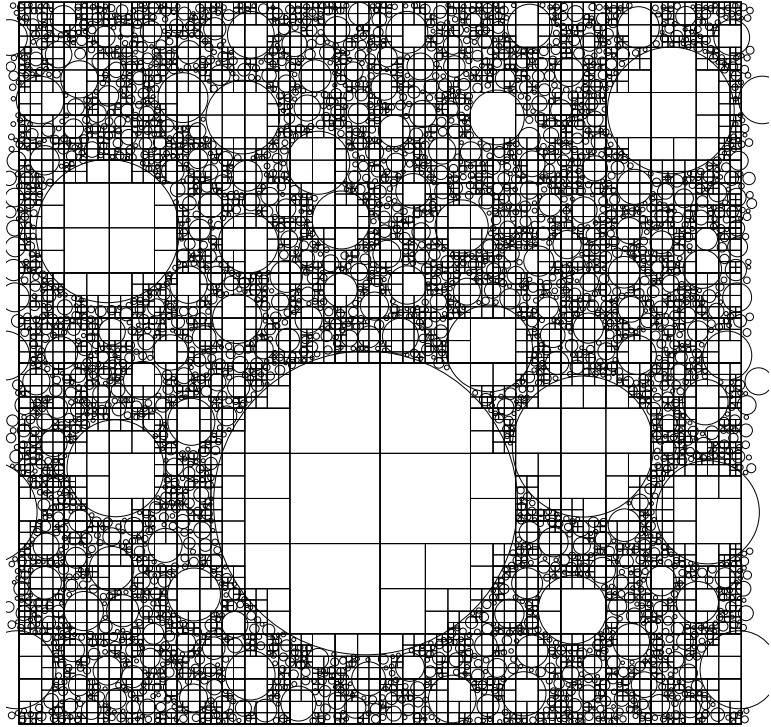}
		\caption{Grid created by the Quadtree}
		\label{fig:quadtree}
	\end{figure}

	\begin{figure}
		\centering
		\includegraphics[width=0.7\textwidth]{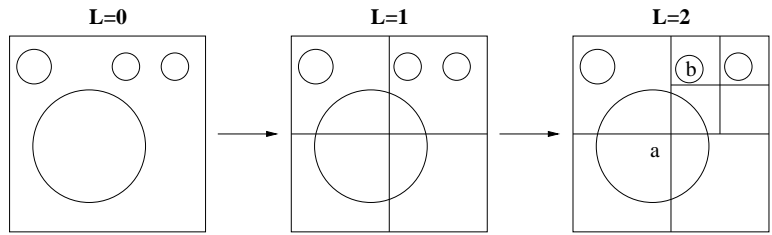}
		\caption{Generating a quadtree in two dimensions.
		Starting from the root cell on level $L=0$, 
		we subdivide the space into $n=4$ subcells which are then on level $L=1$.
		As there are two particles in the upper right corner, this cell is again
		divided into 4 cells.
		}
		\label{fig:level}
	\end{figure}
%	\bild{ebenen.eps}{The principle structure of the quadtree where, starting from the root cell on level 0,	we subdivide the space into $n$ subnodes which are then on level 1.}{fig:level}

	Therefore we use an alternative algorithm, namely the two dimensional quadtree, shown in Fig.
	\ref{fig:quadtree}.
	The quadtree is a	grid with variable cell size.  It is finer where
	there are many small
	particles while it is coarse around big particles. The creation of the grid
	starts with the root cell, which contains the whole system and uses the following
	rule: If a cell contains more than one particle, subdivide it into $n$
	smaller cells and transfer each particle into that new cell where its center
	lies. Continue for the new cells if they contain more than one particle.
	The number of new cells $n$ depends on the dimension $D$ and is given by:
	$n=2^D$.
	
	Fig. \ref{fig:level} shows an example in two dimensions.  Starting from the
	root cell on level $L=0$, we subdivide the space into $n=4$ subcells which
	are then on level $L=1$.  As there are two particles in the upper right
	cell, this cell is again divided into 4 cells.
	For practical reasons we define a maximum depth level $L_{max}$ at which
	we stop subdividing a cell even if it	contains more than one particle. 	
	With $d_t=\frac{l}{2^{L_{max}+1}}$, half
	the size of the smallest cell, a large $L_{max}$ results in a smaller
	Verlet list but we need to update it more often as $d_t$ becomes smaller.
	
	We use the quadtree to create the Verlet list. A contact is added if the
	separation between two particles is smaller than $d_t$, which is half the size
	of the smallest cell.  If a particle travels further than $d_t$ we have to
	regenerate the quadtree and update the Verlet list.
	To create the quadtree and determine the contacts added to the Verlet list,
	we need to know the following lists for each cell: 
	
	\begin{itemize} 
		\item list	$\mathbb{A}$: particles belonging to this cell 
		\item list $\mathbb{B}$: particles overlapping this cell 
		\item list $\mathbb{N}$: neighboring cells
%		\item list $\mathbb{S}$: neighboring cells we will search through when we
%				determine the Verlet list. This list is a subset of $\mathbb{N}$ and
%				prevents visiting cells twice.  
	\end{itemize} 
	
	As the purpose of the last two lists is not evident, we will explain how they
	are used when building the Verlet list.
	First, we compile a
	seperate list of leaves, which are cells without subcells. 
	Now
	recall that all particles are contained in the lists $\mathbb{A}$ of the
	leaves.  If we would follow the linked cell algorithm we would create the
	Verlet list simply by checking the particles of each leaf with those in the
	neighbouring leaves. However, due to the polydispersity of the system we
	would miss several contacts. For instance, in Fig. \ref{fig:level}, the
	contact between the big particle $a$, whose center lies in the lower left
	cell, and the small particle $b$ would be missed.  This problem is solved by
	checking the list $\mathbb{B}$ of the neighboring cells as well. 
	A second difficulty is that the quadtree is an adaptive structure that 
	gives a different neighbor list each time it is created. 
	As the generation on the fly would be too
	costly we store $\mathbb{N}$ during the generation of the quadtree.

%	In the linked cell algorithm $\mathbb{N}$ is not needed because we have a
%	fixed geometry.

%	As we only need to check half
%	of the neighbouring cells $\mathbb{S}$ is built in parallel to $\mathbb{N}$
%	during the quadtree creation. 
%	Using $\mathbb{S}$ avoids the detection of already existing contacts
%	when checking $\mathbb{A}$ of the neighboring leaves.
% 	But a	particle might be listed in $\mathbb{A}$ of one cell and in 
%	$\mathbb{B}$ of another one. The resulting double listing can only be avoided
%	when we check if the contact already exists before adding it to the
%	Verlet list. If it exists it is discarded.

%	Thus we do the following:		For every single leaf
%	$\Delta$ we add all contacts among the particles listed in $\mathbb{A}$ to
%	the Verlet list.  For every cell $\Gamma$ listed in $\mathbb{S}$ of $\Delta$
%	we add the contacts between the particles listed in $\mathbb{A}$ of $\Delta$
%	and $\mathbb{A}$ of $\Gamma$.  Then we add the contacts between the
%	particles in $\mathbb{A}$ of $\Delta$ and $\mathbb{B}$ of $\Gamma$. As a
%	particle might be in $\mathbb{A}$ of one cell and int $\mathbb{B}$ of
%	another cell we check if the contact already exists before adding it to the
%	Verlet list. If it already exists it is discarded.

%% file: Analysis.tex
\section{Method of analysis}
	The shearing is characterized by three parameters, 
	the angle of dilatancy $\Psi$ and the saturation dilatancy $d_s$
	which characterize the movement of the lid
	and the force $F$ needed to maintain the shearing motion.
	%First we will explain these parameters in more detail.

	\subsection{Dilatancy}
		\begin{figure}
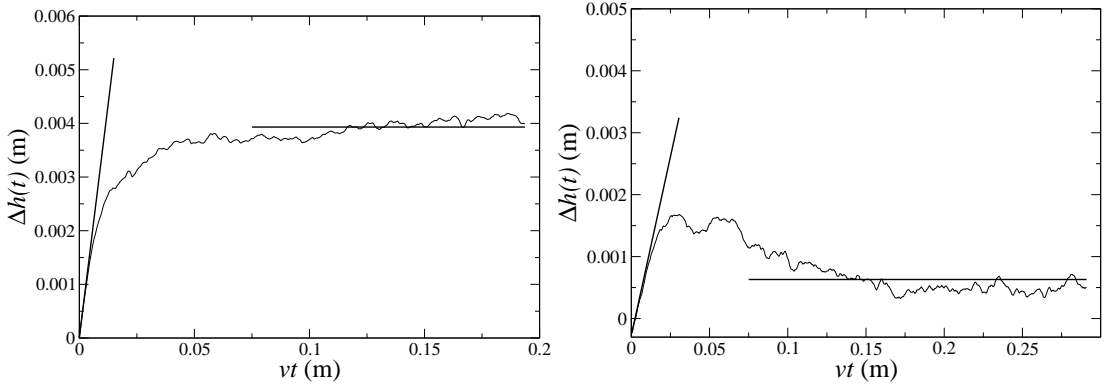

			\subfigure{\includegraphics[width=0.5\textwidth]{Bilder2/winkel_polydisperse.eps}}
			\subfigure{\includegraphics[width=0.5\textwidth]{Bilder/winkel_bidisperse.eps}}
			\caption{\label{fig:1} 
					Expansion $\Delta h(t)$ over the shear distance $s$.
				 	The left picture shows a polydisperse mixture and the right a 
					bidisperse mixture with $R = 1/45$.
					In both cases, $\Phi=0.887$, $v = 1.5 \textrm{m/s}$ and $\Pi=1/45$.
 					For small values of $s$, $\Delta h(t)$ increases almost linearly. 
					For larger values of $s$, $\Delta h(t)$ fluctuates around a 
					saturation value. 
					The straight lines
					show the fits used to obtain $\Psi$ and $d_s$ from $h(t)$.  
					} 
		\end{figure}

		Fig. \ref{fig:1} shows $\Delta h(t)=h(t)-h_0$ over $s=vt$ for a
		polydisperse and a bidisperse system.  The expansion $\Delta h(t)$ tells
		us how far the lid moved from its starting position $h_0$ while the shear
		distance $s$ is the distance the lid moved horizontally. One can identify
		two different regimes. For small values of $s$, $\Delta h(t)$ increases
		almost linearly. For larger values of $s$, $\Delta h(t)$ fluctuates
		around a saturation value. 
		
		One can characterize this behavior by the angle of dilatancy $\Psi$ and
		the saturation dilatancy $d_s$.  Just after the onset of shearing, the
		height	of the lid can be described by 
		
		\begin{equation}
			h(t) = h_0 + vt \tan \Psi.
		\end{equation} 
		
		Therefore, $\Psi$ can be determined by fitting a straight
		line to $h(t)$ for small $t$ and extracting the slope of the line. 
		Specifically, we do a least squares fit for
		the height using the points where $s < 0.002 \textrm{m}$. Thus we make sure
		we only measure the beginning of the shearing. Sometimes, the lid first
		descends before rising. In this case, we begin the fit when the lid is at
		its lowest position, and continue it until $s$ has increased by $0.002 \textrm{m}$.

		Dilatancy is a measure of how much a medium expands when subject
		to shear. Therefore the saturation dilatancy $d_s$ is calculated using 
		\begin{equation}
			d_s = h_s /h_0.
		\end{equation}
		With $h_s$, the average height the lid is assumed to reach for long times
		and $h_0$, the height of the lid before we start shearing.
		In our simulations, it was often difficult to determine $h_s$ because
		it was often not clear that the height had saturated before the
		simulation had ended.
		We obtained values for $h_s$ by taking the average over $\Delta h(t)=h(t) - h_0$ for $t > 0.05$
		seconds. 

	\subsection{Shearing force}
		\begin{figure}
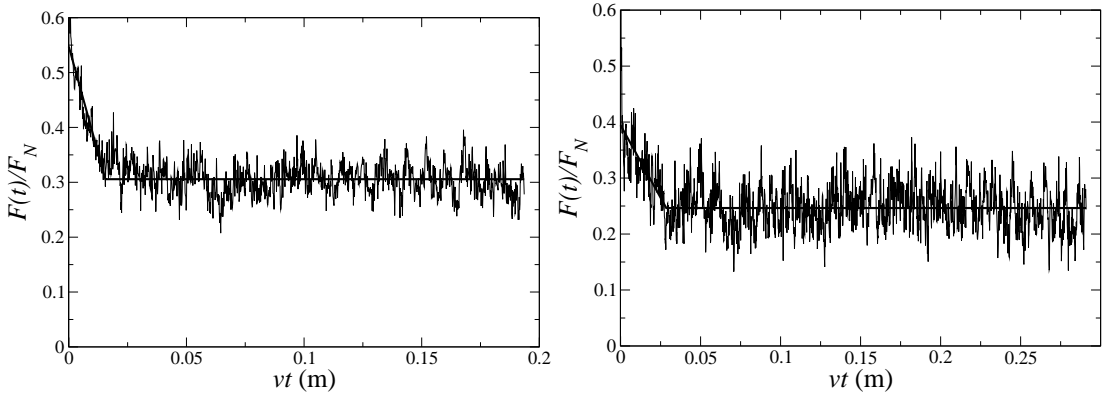

			\subfigure{\includegraphics[width=0.5\textwidth]{Bilder2/force_polydisperse.eps}}
			\subfigure{\includegraphics[width=0.5\textwidth]{Bilder/force_bidisperse.eps}}
			\caption{\label{fig:4}
				The horizontal force $F(t)$ divided by the normal force $F_N$, as
				a function of the shear distance $s$, for the simulations shown in Fig.
				\ref{fig:1}.  
				Again, the polydisperse system is displayed on the left while the right
				picture shows the bidisperse one.
				When the shearing started, the force is high but decreases
				over time until it fluctuates around a saturation force $F_s$. 
				Ten simulations were averaged
				together to obtain these curves. The straight lines show
				the result of fitting the force to the function in Eq.
				\ref{eq:force}.
				}			
		\end{figure}
		
		Fig. \ref{fig:4} shows the horizontal force $F(t)$ divided by the normal
		force $F_N$, as a function of the shear distance $s$, for the
		simulations shown in Fig. \ref{fig:1}.  
		In order to maintain a constant shear velocity $v$ 
		a force $F(t)$ must be exerted on the lid.
		When the shearing started, the force is high but decreases
		over time until it fluctuates around a saturation force $F_s$.
		We are only interested in the resulting saturation force and thus
		need a method to cut off the part of the data where the
		force is still decreasing. This we did by fitting:
%%		This behavior is described by the following formula and the straight
%%		lines in Fig. \ref{fig:4} are the result of the fitting.
		\begin{equation}
			F(t)= \left\{ \begin{array}{ll}
					F_0 + \alpha vt & t<t_*\\
					F_s & t\ge t_*\\
												\end{array} \right. 
			\label{eq:force}
		\end{equation}
		Here $F_0$ is the force necessary to start the shearing with velocity $v$
		and $\alpha$ is the slope telling us how fast the shearing force approaches
		its saturation value which it reaches at time $t_*$.
		The three parameters $F_0$,
		$\alpha$, and $F_s$ are extracted by fitting the observed force to 
		Eq. \ref{eq:force}. 
		%The time $t_*$ can then be calculated by $t_* = (F_0 - F_s )/(\alpha v)$. 
		Typically $F_s\approx \mu F_N$ and $F_0$ is at the maximum twice as big as $F_s$ while
		$\alpha\approx 1000-25000 \textrm{N/m}$.

%% file: Results.tex
\section{Results}
	We first compare the shearing behavior of bidisperse and polydisperse mixtures
	and in the second part we will investigate how the introduction of point-like particles
	changes this behavior for polydisperse mixtures.

	\subsection{Bidisperse and polydisperse mixtures}
		We investigated two bidisperse mixtures and a 
		polydisperse one. 
		In the polydisperse
		mixture, the sizes are distributed according to Eq. \ref{eq:powerlaw}
		with $b = 3.5$. For the bidisperse mixtures we have $R =1/45$ 
		and $R = 1/60$ [see Eq. \ref{eq:bidispers}].  In both
		bidisperse mixtures, $\Pi = 10$ ($r_{min}=0.1 \textrm{cm}$, $r_{max}=1\textrm{cm}$).  
		The simulations were done with
		either 575 (polydisperse), 690 ($R = 1/45$), or 732 ($R = 1/60$)
		particles. 

		We examined each mixture at two or three initial densities. All mixtures
		were studied at $\Psi=0.887$ and $0.876$. In addition the bidisperse
		mixtures were examined at $\Psi=0.911$. For each mixture-density pair,
		ten different samples were prepared and the shearing velocity was set to
		three different values: $v=0.5,1.5,4.5 \textrm{m/s}$.  The time series from
		each group of ten simulations were averaged together to obtain the
		shearing parameters.  

		\begin{figure}
			\centering
			\includegraphics[width=0.9\textwidth]{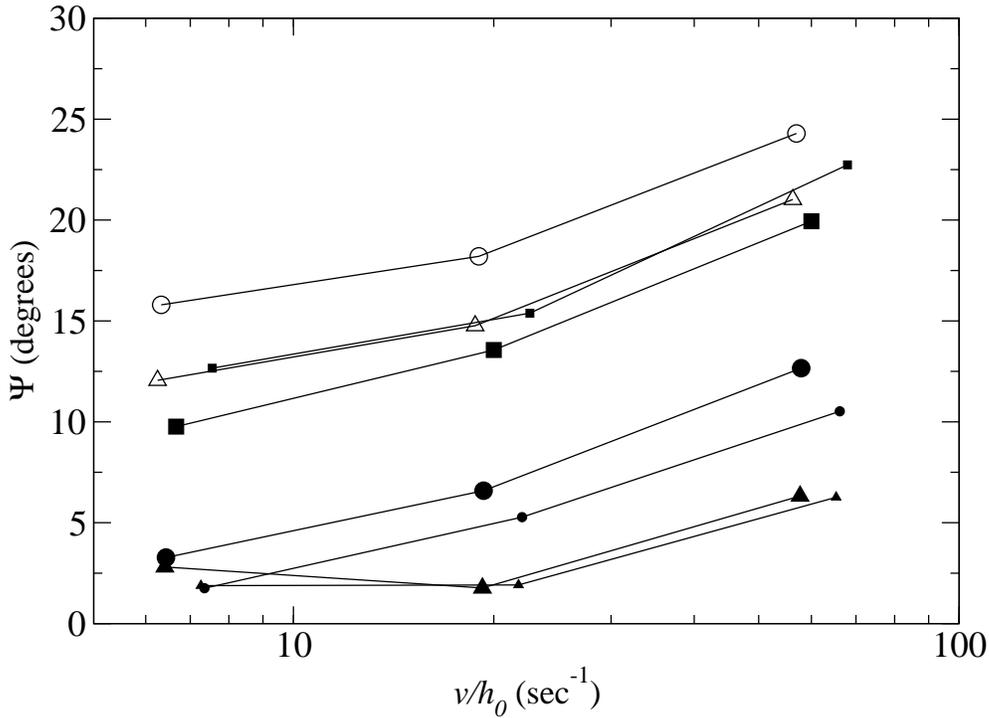}
			\caption{\label{fig:2}
				Dilatancy angles $\Psi$ for bidisperse and polydisperse mixtures,
				as a function of the initial shear rate $v/h0$ . The empty symbols
				correspond to polydisperse mixtures; the large, filled symbols
				correspond to a bidisperse mixture with $R = 1/45$, and the small,
				filled symbols to a bidisperse mixture with $R = 1/60$. The squares
				indicate results for an initial solid fraction $\Phi_0 = 0.911$,
				the circles $\Phi_0 = 0.887$ and the triangles $\Phi_0=0.876$. The
				highest density ($\Phi_0 = 0.911$) could only be obtained with the
				bidisperse mixtures.
				The angle of dilatancy increases with shearing velocity and 
				density but is roughly three times smaller for bidisperse
				mixtures than for polydisperse ones. 
				}			
		\end{figure}
		
		In Fig. \ref{fig:2}, we show the angle of dilatancy $\Psi$ over the initial
		shear rate $v/h_0$ for the different
		simulations we performed. Some trends can be seen. 
		The angle of dilatancy increases with shearing velocity and density
		but is roughly three times smaller for bidisperse
		mixtures than for polydisperse ones. On the
		other hand, at the maximum density ( $\Phi_0 = 0.911$), the bidisperse mixture's
		angle approaches those of the polydisperse mixture. Note that this
		density could not be obtained for the polydisperse mixture. 

		\begin{figure}
			\centering
			\includegraphics[width=0.9\textwidth]{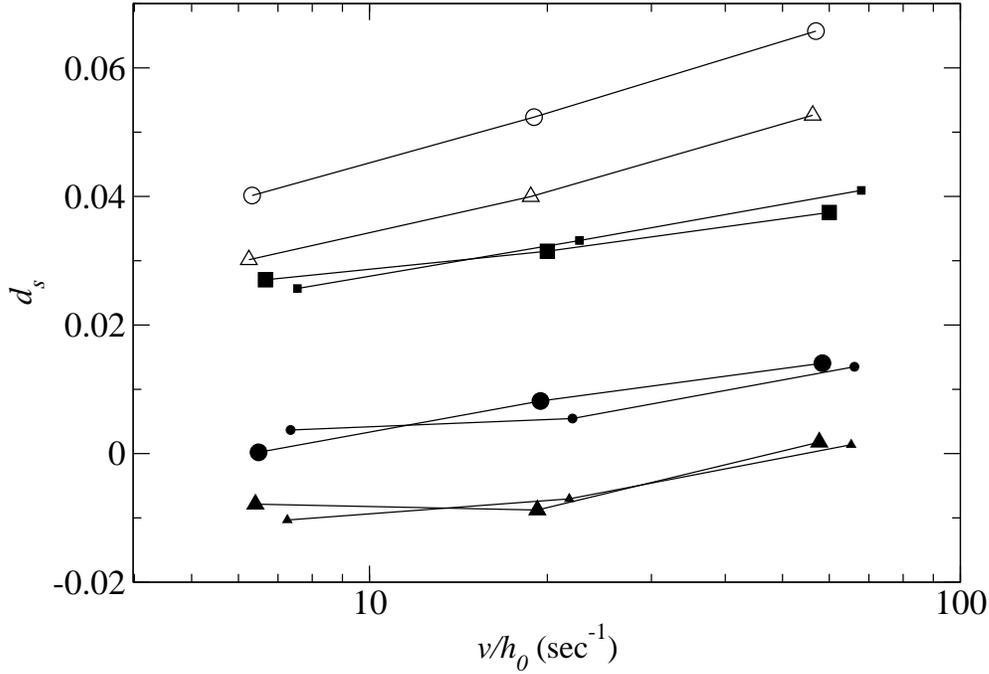}
			\caption{\label{fig:3}
				Saturation dilatancy $d_s$ for polydisperse and bidisperse particles.
				The symbols are the same as those used in Fig \ref{fig:2}.
				As for the angle of dilatancy the saturation dilatancy increases 
				with velocity and density and 
				is for polydisperse particles greater then for bidisperse mixtures.
				Note the negative dilation for the bidisperse system displayed by
				the large filled triangles.
				}		
		\end{figure}

		The saturation dilatancy for the systems we studied is shown in
		Fig. \ref{fig:3}. Here we plot $d_s$ over the initial shear rate.
		As for the angle of dilatancy the saturation dilatancy increases 
		with velocity and density and 
		is for polydisperse particles greater then for bidisperse mixtures. 
		An important difference is the fact that for the lowest initial density only
		bidisperse mixtures exhibit negative dilation.
		This
		means that the height $h_s$ at the end of shearing is lower than $h_0$ 
		at the
		beginning. This occurs because the small particles do not fill all the
		spaces between the large particles during the preparation of the sample.
		When the shearing begins, there can be large voids between the big
		particles. As the shearing proceeds, the large particles move relative to
		one another, the voids are opened up, and quickly filled with small
		particles. The voids never re-form, leading to a permanent decrease in
		the height of the lid.

		\begin{figure}
			\centering
			\includegraphics[width=0.9\textwidth]{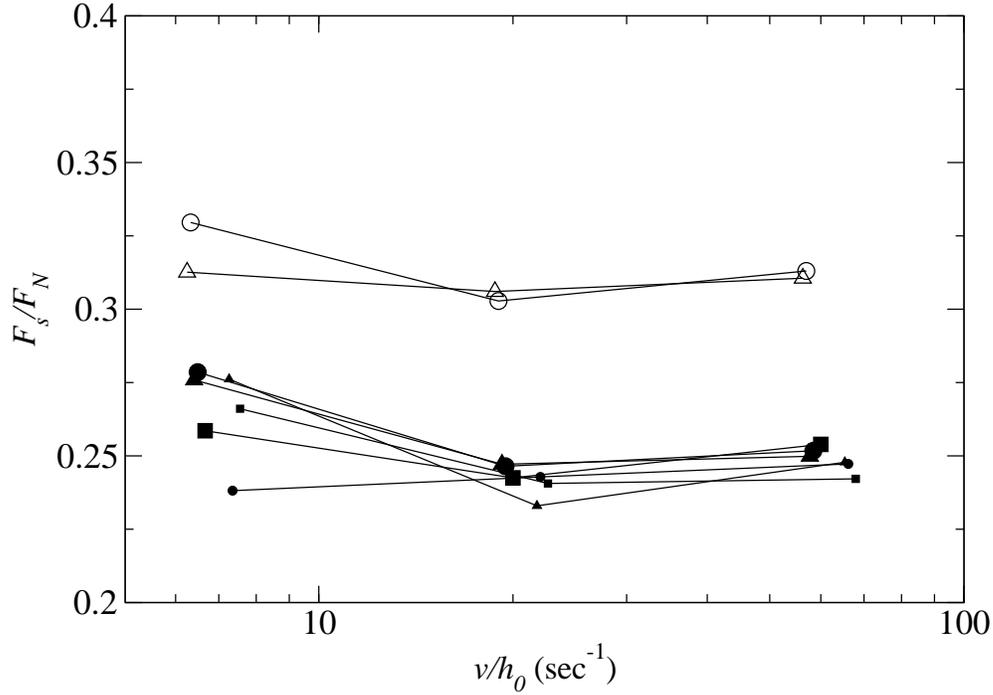}
			\caption{\label{fig:5}
				The saturation value of the force $F_s$ , divided by the imposed
				normal force $F_N$, over the initial shear rate for polydisperse and bidisperse particles. The
				symbols are the same as those used in Fig. \ref{fig:2}. Note that
				all values are close to the Coulomb friction ratio $\mu = 0.3$ used
				in the simulations.
				The polydisperse particles have a force
				that is roughly 30\% higher than the bidisperse ones. The force is
				roughly independent of the initial density. Surprisingly, it
				decreases slightly with velocity, at least between $v = 0.05$ and $v =
				0.15$.
				}			
		\end{figure}

		In Fig. \ref{fig:5}, we show the saturation force $F_s$ divided by $F_N$ over the initial
		shear rate for the
		different series of simulations. 
		The polydisperse particles have a force
		that is roughly 30\% higher than the bidisperse ones. The force is
		roughly independent of the initial density. Surprisingly, it
		decreases slightly with velocity, at least between $v = 0.5 \textrm{m/s}$ and $v =
		1.5 \textrm{m/s}$. This differs from other cases, where the force is always observed to
		increase with shearing velocity \cite{Physica_A_217_261}. However, that
		work concerns flow of approximately monodisperse polygons, whereas we
		have studied disks. The decrease in $F_s$ can be understood as a
		consequence of dilatancy. At higher velocities, dilatancy increases and
		thus making it easier to shear. 
		
	%	This effect will be absent when the
	%	vertical position of the wall is fixed.
	%	\textbf{as ....}

\subsection{Point-like Particles}
	In this section we systematically 
	study how the shearing parameters for a polydisperse
	mixture change when point-like particles are added.
	Again, the size distribution in the polydisperse
	mixture follows Eq. \ref{eq:powerlaw}
	with an exponent $b=3.5$.

	For the point-like particles there are three parameters we can change.  The
	first two are the
	strength $k$ and the range of the repulsive force $d_r$ in the repulsive
	potential in Eq. \ref{eq:repulsivepotential}.  The third parameter 
	$A=\frac{N(\textrm{point-like})}{N(\textrm{grains})}$ gives the
	number of point-like particles, divided by the number of non point-like
	particles. 
	We set the following values for these parameters:
	$A=1,2$, $d_r=\frac{1}{5}r_{max}, \frac{2}{5}r_{max}$ and $k=2,5,10\times 10^{-3}N$ 
	For all simulations we set $\Phi=0.882$ and the shearing velocity to $v=1.5 \textrm{m/s}$.
	In both mixtures we have $\Pi=10$ and $r_{min}=0.11 \textrm{cm}$, $r_{max}=1.1 \textrm{cm}$.
	For each set of these three parameters, ten simulations with different 
	initial configurations were done and averaged together. For some systems we even set
	the strength of the potential to $k=20, 50, 100\times 10^{-3}N$.

\begin{figure}
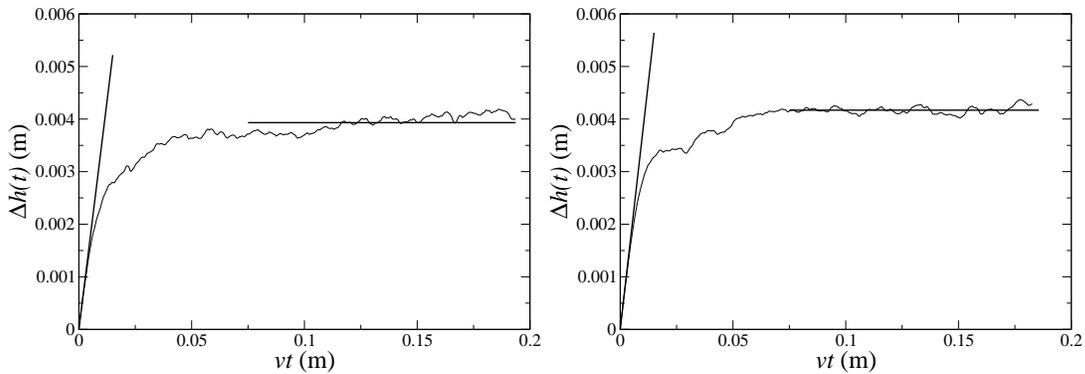

		\begin{center}
		\includegraphics[width=0.49\textwidth]{Bilder2/winkel_polydisperse.eps}
		\includegraphics[width=0.49\textwidth]{Bilder2/winkel_punkt.eps}
		\end{center}
		\caption{\label{winkel_example} 
			Expansion $\Delta h(t)$ over the shear distance $s$. The left picture
			shows the polydisperse mixture, the right picture shows
			the same mixture, with one
			point-like particle added for every grain ($A=1$).  
			In both cases, $\Phi_0=0.882$ and $v=1.5 \textrm{m/s}$.
			The repulsive
			potential has a strength of $k=10^{-3}N$ and interaction distance
			$d=0.22\textrm{cm}$ ($1/5$ the radius of the largest particles). 
			Ten simulations were averaged together to obtain these
			curves.  The straight lines show the fits used to obtain $\Psi$ and
			$d_s$.
 			One can see that the presence of the point-like particle changes
			the curve very little.
			}
		\end{figure}

	In Fig. \ref{winkel_example} we show the 
	expansion $\Delta h(t)$ over the shear distance $s$ for a polydisperse
	and the same polydisperse system
	with one	point-like particle added for every grain ($A=1$).  
	The straight lines show the fits used to obtain $\Psi$ and
	$d_s$. One can see that the presence of the point-like particle changes
	the curve very little.

	\begin{figure}
		\begin{center}
		\includegraphics[width=0.8\textwidth]{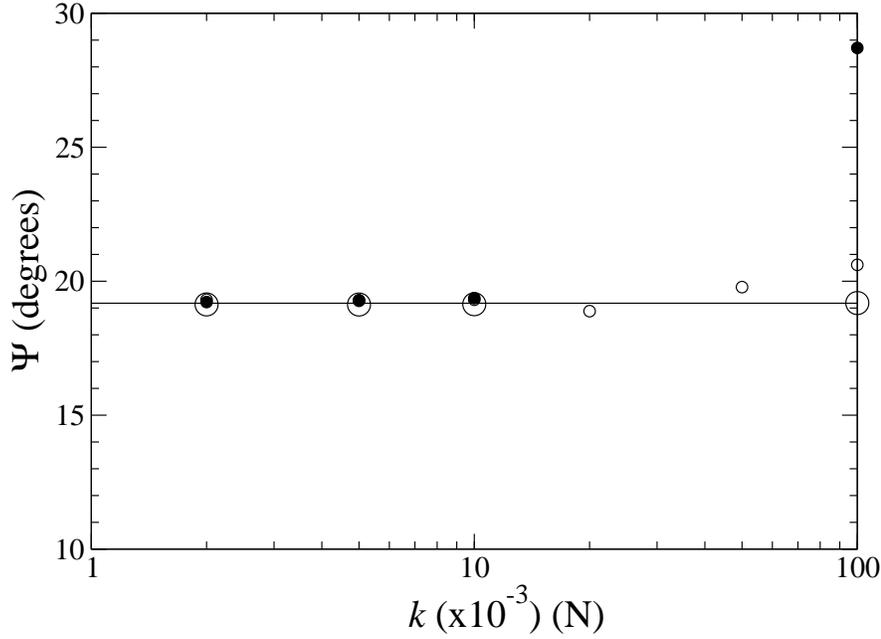}
		\end{center}
		\caption{\label{winkel_results}
			Dilatancy angles $\Psi$ for polydisperse mixtures with and without
			point-like particles, as a function of the the strength of the
			repulsive potential $k$.  The horizontal line shows the value of
			$\Psi$ obtained when there are no point-like particles.  The small
			circles correspond to repulsive particles with a small interaction
			radius ($1/5$ the radius of the largest particle) and large circles
			correspond to a large interaction radius ($2/5$ of the largest
			particle).  The empty symbols represent simulations with equal numbers
			of repulsive particles and grains ($A=1$); the filled symbols
			represent simulations when the number of repulsive particles has been
			doubled ($A=2$).  In all cases, the density is $0.882$ and the
			shearing velocity is $1.5 \textrm{m/s}$.
			The addition of point-like
			particles does not cause much change. 
			The observed angles vary by about $1$
			degree from the value found without point-like particles.
			} 
		\end{figure}

	In Fig.~\ref{winkel_results}, we plot $\Psi$ over $k$ 
	for the
	different types and numbers of point-like particles.  One data point stands
	out from the rest: ($\Psi\approx29^{\circ}$, $k=100\times10^{-3} N$, $A=2$).
	For the moment, we will exclude it from our discussion and treat it in a
	special section, Sec.~\ref{exceptionalpt}.
	Except for this one series of simulations, the addition of point-like
	particles does not cause much change.  The observed angles vary by about one
	degree from the value found without point-like particles.
	Compared with the changes we discussed in the previous section the changes due
	to the point-like particles are one magnitude smaller then when changing
	the shear velocity or the initial density.
	
	A very surprising feature of the shearing behavior extracted from 
	Fig.~\ref{winkel_results} is that point-like
	particles with a large distance of interaction cause less change than
	particles with a small distance.  (Compare the large and small empty circles
	at $k=100\times10^{-3} N$.)  

	\begin{figure}
		\begin{center}
		\includegraphics[width=0.8\textwidth]{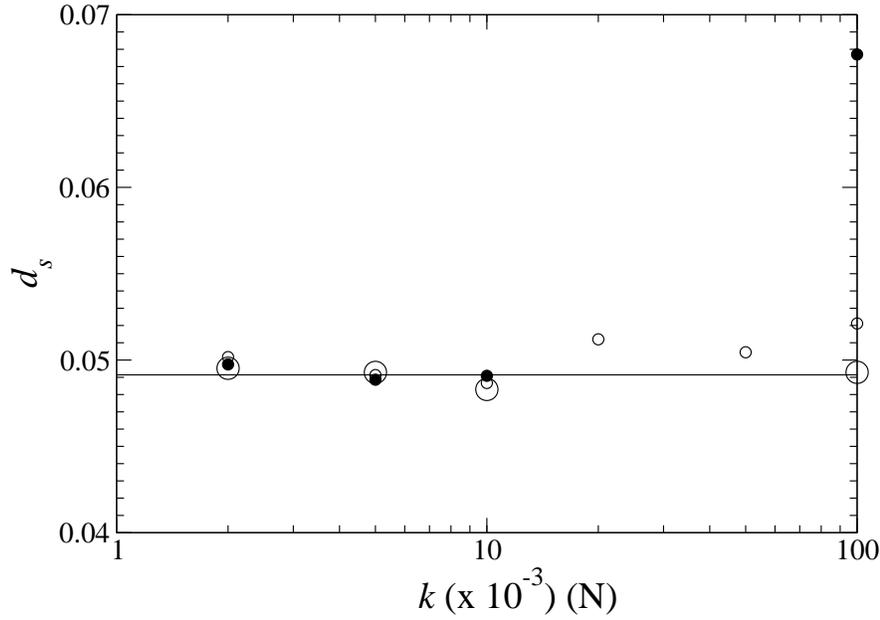}
		\end{center}
		\caption{\label{dilation_results}
			Saturation dilatancy $d_s$ for mixtures with different concentrations
			of point-like particles.  The symbols are the same as those used in
			Fig.~\ref{winkel_results}. The presence of point-like particles changes the
			behavior of the mixture very little.
			} 
	\end{figure}

\begin{figure}
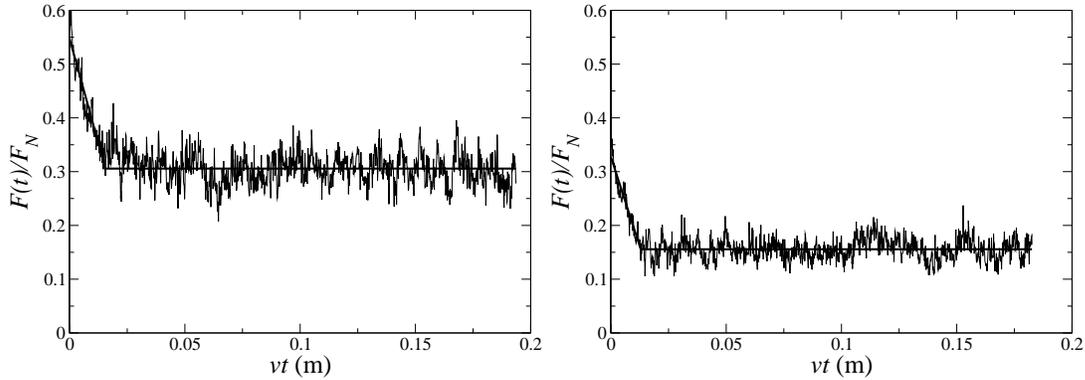

		\begin{center}
		\includegraphics[width=0.49\textwidth]{Bilder2/force_polydisperse.eps}
		\includegraphics[width=0.49\textwidth]{Bilder2/force_punkt.eps}
		\end{center}
		\caption{\label{force_example} 
			The horizontal force $F(t)$ divided by the normal force $F_N$, as a
			function of the shear distance $s$, for the simulations shown in
			Fig.~\ref{winkel_example}. 
			The left picture shows the polydisperse mixture
			and the right picture shows the same 
			simulation with point-like particles ($A=1$ and $k=100\times10^{-3} N$.
		  In both cases, $\Phi_0=0.882$ and $v=1.5 \textrm{m/s}$.
			Ten simulations were averaged together to obtain these curves.  The
			straight lines show the result of fitting the force to the function in
%			Eq.~(\ref{force_theory}).
Eq.~(\ref{eq:force}).
			} 
	\end{figure}

	The saturation dilatancy $d_s$ over the strength $k$ of the potential
	for the systems studied is shown in
	Fig.~\ref{dilation_results}.  The data are very similar to those discussed above.
	The simulations with
	$k=100\times10^{-3}N$ and $A=2$ are widely separated from all the others.
	Except for this one data point, the presence of point-like particles changes
	the behavior very little.  When point-like particles are added, the dilatancy
	changes by at most $0.003$.
	The changes due to the change in density or shear rate, observed in the previous section were seven time larger.	
	In Fig. \ref{force_example} we show the horizontal force $F(t)$ 
	divided by the normal force $F_N$, as a
	function of the shear distance $s$, for the simulations shown in
	Fig.~\ref{winkel_example}. 

\begin{figure}
\begin{center}
\includegraphics[width=0.8\textwidth]{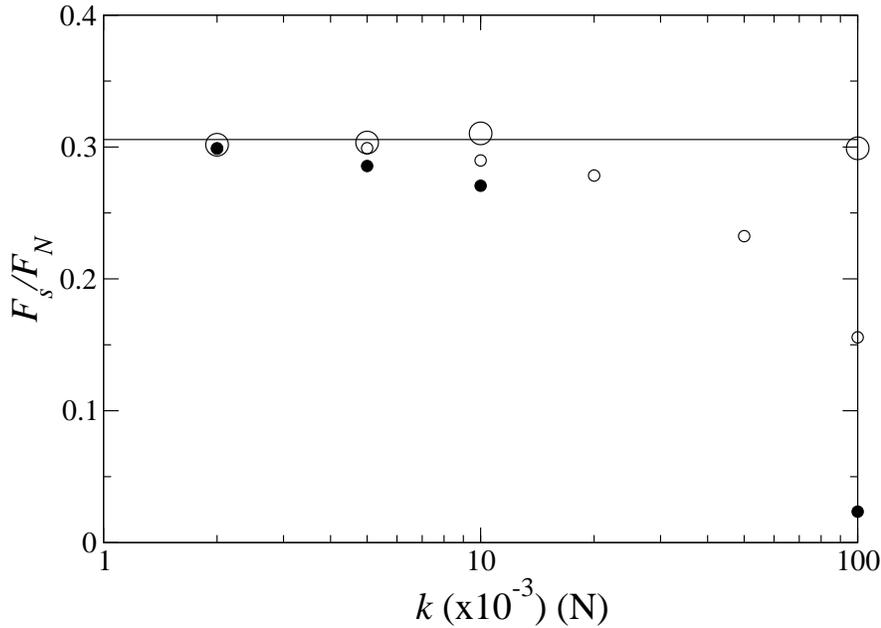}
\end{center}
\caption{\label{force_results}
The saturation value of the force $F_s$, divided by the imposed
normal force $F_N$, for systems with different concentrations of
point-like particles.
The symbols are the same as those used in Fig.~\ref{winkel_results}.
Note that $F/F_N\approx\mu$ when $k$ is small.
When short-range point-like
particles are added to the polydisperse mixture (with $d_r=r_\mathrm{max}/5$, the small circles in the figure), the force decreases
substantially.  At large $k$, the force is reduced to half of its original
value, or very nearly removed, depending on how many point-like particles
are added.  
}
\end{figure}

In Fig.~\ref{force_results}, we show the saturation force $F_s$ over $F_N$ 
when the lid fluctuates about its saturation height $h_s$
for the different series of simulations. This time, the point-like
particles change the behavior of the mixture.  When short-range point-like
particles are added (with $d_r=r_\mathrm{max}/5$, the small circles in the figure), the force decreases
substantially.  At large $k$, the force is reduced to half of its original
value, or very nearly removed, depending on how many point-like particles
are added.  

Surprisingly, adding the long-range particles does not
reduce the force at all.
This continues the general trend observed in the previous discussion,
where we saw that the particles with a large interaction distance did not
have much effect.  One possible reason for this is that when $d_r$ is small,
the grains feel only those point-like particles that occupy the neighboring
pore spaces.  The force exerted on the grains is thus tightly connected to
the geometry of the surrounding particles, and this force is such that it
reduces the friction between the grains.  When $d_r$ is large, grains
interact with point-like particles in many different regions, and the resulting 
forces are no longer so closely related to the geometry of the neighboring
particles.

\subsubsection{Behavior at large $k$ and $A$}
\label{exceptionalpt}

In Fig.~\ref{winkel_results}, we observed for large $k$ and $A$
a very high angle of dilatancy (nearly $30^{\circ}$ compared with 
all the other points near $20^{\circ}$.), and in Fig. 
\ref{dilation_results}
a very high
saturation dilatancy (roughly $40\%$ more than any other point).  
In addition, the force needed to shear this mixture is 
very low -- only $10\%$ of the mixture without point-like particles.
This last result suggests that the point-like particles are carrying a significant
fraction of the weight of the lid.  This would explain the very high dilatancy.
Note that all the samples are prepared by compressing mixtures with $k=10^{-3}N$.
Then, when the shearing starts, $k$ is set to its final value.  If $k$ is
very large and the point-like particles are numerous, the mixture will expand
not due to shearing, but simply because the point-like particles push against
each other with enough force to lift up the lid.

This explanation has been confirmed by simulations of unsheared systems.
The system is prepared as before, but is not sheared.  When $k=100\times10^{-3}N$
and $A=2$, we observe a substantial dilation ($d_s\approx0.28$)
due only to the repulsive potential
of the point-like particle.  On the other hand,
when $k=100\times10^{-3}N$ and $A=1$, no such dilation is observed.  We conclude,
therefore, that the point at $k=100\times10^{-3}N$ and $A=2$ is in a different
regime from the other points.

\subsubsection{Dependence on polydispersity}

All of the above results were obtained with a polydisperse mixture described
by a power law exponent $b=3.5$ [see (\ref{eq:powerlaw})].  We also tried
mixtures with $b=1.3$.  Point-like particles with $d_r=0.022$ and
$2\times10^{-3} N\le k \le 100\times10^{-3} N$ were added.
No significant change in the dilatancy or force is observed, even for
the largest values of $k$.
This may be because there are fewer small grains,
and the pore spaces are much larger.  The point-like particles can then
stay in the middle of this pore spaces and interact only weakly with the grains.

%% file: Conclusion.tex
\section{Conclusion}
The findings of this study can be summarized by saying that the
polydisperse mixtures show stronger dilatancy and a greater resistance to
shearing than the bidisperse mixtures. At constant density, the angle of
dilatancy, the saturation dilatancy, and the force needed to maintain the
shearing were all greater for polydisperse particles. However, this simple
conclusion is complicated by the fact that higher densities were easier to
obtain with bidisperse mixtures. When bidisperse mixtures are very dense, their
angle of dilatancy and saturation dilatancy is similar to polydisperse systems
at lower densities (although the shearing force remains significantly smaller).

Adding repulsive particles to a sheared
polydisperse mixture of
grains changes the
kinematic behavior of the mixture very little,
but the dynamic behavior shows a reduction in the forces.  By "kinematic"
we mean those properties that concern the movement of the mixture --
the angle of dilatancy and the saturation dilatancy.  By "dynamic"
behavior, we mean the force necessary to maintain a fixed shearing
velocity.
This finding is complicated by two additional observations.  First, 
particles with a large interaction distance cause little change,
in spite of exerting larger forces.  The second observation is that it
is possible to get dramatic changes in the kinematic behavior when
there are many point-like particles with strong repulsive
forces. 

In general we can say that the point-like particles lead to a lubrication
effect which reduces the force necessary to shear the system. But one has to
be careful not to add too many point-like particles. If the
number of point-like particles becomes too large, they will 
build a network that carries most of the load and leads to a strong dilation
after the initialization.
For future work it might be interesting to see how
the lubrication effect changes when using different normal forces on the lid.

%\section{Acknowledgement}
\ack
We thank Dr. Dieter Distler from BASF for interesting and fruitful 
	discussions and support.

%% file: paper.bbl
\begin{thebibliography}{10}

\bibitem{Phys_Rev_E_55_4_4720}
S.Luding.
\newblock Stress distribution in static two-dimensional granular model media in
  the absence of friction.
\newblock {\em Physical Review E}, 55(4):4720--4729, 1997.

\bibitem{Phys_Rev_E_54_1_861}
Farhang Radjai and Lothar Brendel.
\newblock Nonsmoothness, indeterminacy, and friction in two-dimensional arrays
  of rigid particles.
\newblock {\em Physical Review E}, 54(1):861--873, 1996.

\bibitem{Phys_Rev_Lett_77_2_274}
Farhang Radjai, Michael Jean, Jean-Jaques Moreau, and Stéphane Roux.
\newblock Force {D}istribution in {D}ense {T}wo-{D}imensional {G}ranular
  {S}ystems.
\newblock {\em Physical Review Letters}, 77(2):274--277, 1996.

\bibitem{Moreau}
J.J. Moreau.
\newblock {\em Lecture Notes in Applied and Computational Mechanics}, chapter
  Novel Approaches in Civil Engineering.
\newblock 2004.

\bibitem{Phys_Rev_Lett_77_15_3110}
Brian Miller, Corey O'Hern, and R.~P. Behringer.
\newblock {S}tress {F}luctuations for {C}ontinuously {S}heared {G}ranular
  {M}aterials.
\newblock {\em Physical Review Letters}, 77(15):3110--3113, 1996.

\bibitem{Chaos_9_3_559}
Daniel~W. Howell and R.~P. Behringer.
\newblock Fluctuations in granular media.
\newblock {\em Chaos}, 9(3):559--572, 1999.

\bibitem{Philos_Mag_20_469}
O.~Reynolds.
\newblock On the dilatancy of media composed of rigid particles in contact.
\newblock {\em Philo. Mag.}, 20:469, 1885.

\bibitem{Phys_Rev_E_59_5_5881}
J.~Géminard, W.~Losert, and J.~Gollub.
\newblock Frictional mechanics of wet granular material.
\newblock {\em Physical Review E}, 59(5):5881--5890, 1999.

\bibitem{Phys_Rev_Lett_79_5_949}
S.~Nasuno, A.~Kudrolli, and J.~P. Gollub.
\newblock Friction in {G}ranular {L}ayers: {H}ysteresis and {P}recursors.
\newblock {\em Physical Review Letters}, 79(5):949--952, 1997.

\bibitem{Phys_Rev_E_59_1_739}
C.~T. Veje, Daniel~W. Howell, and R.~P. Behringer.
\newblock {K}inematics of a two-dimensional granular {C}ouette experiment at
  the transition to shearing.
\newblock {\em Physical Review E}, 59(1):739--745, 1999.

\bibitem{Phys_Rev_Lett_67_13_1751}
P.~Thompson and G.~Grest.
\newblock Granular {F}low: {F}riction and the {D}ilatancy {T}ransition.
\newblock {\em Physical Review Letters}, 67(13):1751--1754, 1991.

\bibitem{Physica_A_217_261}
HJ~Tillemans and H.J. Herrmann.
\newblock Simulation deformations of granular solids under shear.
\newblock {\em Physica A}, 217:261--288, 1995.

\bibitem{Eur_Phys_J_E_2_2_181}
F.~Lacombe, S.~Zapperi, and H.J.Herrmann.
\newblock Dilatancy and friction in sheared granular media.
\newblock {\em Eur. Phys. J. E}, 2(2):181--189, 2000.

\bibitem{Eur_Phys_J_E_2002_10160_7}
M.Lätzel, S.Luding, H.J. Herrmann, D.W. Howell, and R.P. Behringer.
\newblock Comparing simulation and experiment of a 2d granular {C}ouette shear
  device.
\newblock {\em Eur. Phys. J. E}, 11:325--333, 2002.

\bibitem{condmat_0407100}
James~F. Lutsko.
\newblock The rheology of dense, polydisperse granular fluids under shear.
\newblock {\em Condmat}, (0407100), 2004.

\bibitem{Tildesley}
M.P.Allen and D.J.Tildesley.
\newblock {\em Computer {S}imulation of {L}iquids}.
\newblock Oxford University Press, Oxford, 1987.

\end{thebibliography}
